\documentclass[]{spie}  

\usepackage{amsmath,amsfonts,amssymb}
\usepackage{graphicx}
\usepackage{subfigure}
\usepackage{multirow}
\usepackage[colorlinks=true, allcolors=blue]{hyperref}
\usepackage{booktabs}
\usepackage{xcolor}
\usepackage{setspace}

\title{Ambient-Pix2PixGAN for Translating Medical Images from Noisy Data}

\author[a]{Wentao Chen}
\author[b]{Xichen Xu}
\author[c]{Jie Luo}
\author[a,b]{Weimin Zhou}
\affil[a]{University of Michigan-Shanghai Jiao Tong University Joint Institute, Shanghai Jiao Tong University, Shanghai, 200240, China}
\affil[b]{Global Institute of Future Technology, Shanghai Jiao Tong University, Shanghai, 200240, China}
\affil[c]{School of Biomedical Engineering, Shanghai Jiao Tong University, Shanghai, 200240, China}

\authorinfo{Further author information: (Send correspondence to Weimin Zhou.)\\Weimin Zhou: E-mail: weimin.zhou@sjtu.edu.cn}

\begin{document} 
\maketitle

\begin{abstract}
Image-to-image translation is a common task in computer vision and has been rapidly increasing the impact on the field of medical imaging. Deep learning-based methods that employ conditional generative adversarial networks (cGANs), such as Pix2PixGAN, have been extensively explored to perform image-to-image translation tasks. However, when noisy medical image data are considered, such methods cannot be directly applied to produce clean images. Recently, an augmented GAN architecture named AmbientGAN has been proposed that can be trained on noisy measurement data to synthesize high-quality clean medical images. Inspired by AmbientGAN, in this work, we propose a new cGAN architecture, Ambient-Pix2PixGAN, for performing medical image-to-image translation tasks by use of noisy measurement data. Numerical studies that consider MRI-to-PET translation are conducted. Both traditional image quality metrics and task-based image quality metrics are employed to assess the proposed Ambient-Pix2PixGAN. It is demonstrated that our proposed Ambient-Pix2PixGAN can be successfully trained on noisy measurement data to produce high-quality translated images in target imaging modality.
\end{abstract}


\section{PURPOSE}
Image-to-image translation is a common computer vision task that aims to map images in one domain (source domain) to images in another domain (target domain). Medical image translation has been rapidly increasing the impact on the field of medical imaging \cite{armanious2020medgan}. For example, MR-to-CT translation was conducted for radiotherapy treatment planning \cite{yang2020unsupervised}; MR-to-PET translation was performed for establishing a group of pseudo-normal PET images to provide healthy control for localizing the epileptic focus \cite{hu2021bidirectional};
Low-dose to high-dose CT image translation was investigated for low-dose CT image denoising \cite{yang2018low}.
Moreover, image-to-image translation may also be useful for establishing stochastic object models (SOMs) that produce object properties in missing imaging modalities. Such SOMs are a critical component for the objective assessment of image quality \cite{zhou2019approximating, zhou2020approximating,li2021hybrid ,zhou2023Ideal,granstedt2023approximating}.  

Deep learning-based methods that employ conditional generative adversarial network (cGAN) hold great promise to perform image-to-image translation tasks for medical image analysis \cite{dar2019image}. A cGAN, such as Pix2PixGAN \cite{isola2017image}, comprises a generator and a discriminator that are both represented by neural networks. In a cGAN training process, a discriminator is trained to capture the target distribution information for guiding the generator to perform a mapping from the source image domain to the target image domain. The generator is trained to produce target images that are not only indistinguishable from the discriminator but also share similar structure as the input source image. However, medical imaging systems acquire measurement data that are noisy and/or incomplete. Therefore, training cGAN directly on noisy medical imaging measurements may lead to degraded image quality.

Zhou \emph{et al.} investigated the ability of AmbientGANs \cite{bora2018ambientgan} to establish SOMs by use of noisy measurement data \cite{zhou2022learning}. In AmbientGANs, the measurement process was included in the GAN architecture to simulate the measurement data corresponding to the generator-produced ``fake'' objects, and the discriminator is trained to distinguish between the simulated measurement data and the ground truth measurement data. The generator is subsequently trained against the discriminator to learn the underlying distribution of the objects to-be-imaged. 

Inspired by AmbientGAN, in this work, we propose a novel paired image-to-image translation method Ambient-Pix2PixGAN that can act on noisy medical image data. Numerical studies that considered MR-to-PET translation are conducted to investigate the ability of our proposed method to perform image-to-image translation tasks by the use of noisy medical image data. Both traditional image quality metrics and task-based image quality metrics are evaluated to assess the trained models. It is demonstrated that the proposed Ambient-Pix2PixGAN method can be successfully trained on noisy medical image data to produce high-quality translated images.

\section{METHODS}
\subsection{Pix2PixGAN} 
Isola \emph{et al.} \cite{isola2017image} proposed a conditional GAN method named Pix2PixGAN to perform paired image-to-image translation tasks. A so-called Pix2PixGAN comprises a generator and discriminator, both are represented by deep neural networks. The generator is trained against the discriminator through an adversarial process to produce images that are indistinguishable to the training images in the target domain. Moreover, in the training, the $L1$ distance between the ground-truth images and the generator-produced images is minimized to ensure that the produced images are close to the ground truth images. More details about the Pix2PixGAN can be found in the publication \cite{isola2017image}. However, Pix2PixGAN that is directly applied to medical image data that are contaminated by measurement noise is not able to produce high-quality clean images, which may subsequently affect downstrem medical imaging tasks. To address this issue, in this work, we develop a novel Ambient-Pix2PixGAN method that can be trained on noisy medical image data for producing clean translated images in the target imaging modality.
\subsection{Ambient-Pix2PixGAN} 
To train a generative model by use of noisy measurement data, Bora \emph{et al.} \cite{bora2018ambientgan} proposed an augmented GAN architecture named AmbientGAN. In AmbientGAN, a measurement process was included to simulate the measurement data corresponding to the generator-produced objects. More recently, Zhou \emph{et al.} \cite{zhou2020learning, zhou2022learning} investigated the use of AmbientGANs to establish stochastic object models from measurement data. Inspired by these works, we propose an Ambient-Pix2PixGAN for performing medical image translation tasks by use of noisy measurement data. The proposed Ambient-Pix2PixGAN architecture is illustrated in Fig. \ref{illustration}.
\begin{figure}[!ht]
    \centering
    \includegraphics[width=\textwidth]{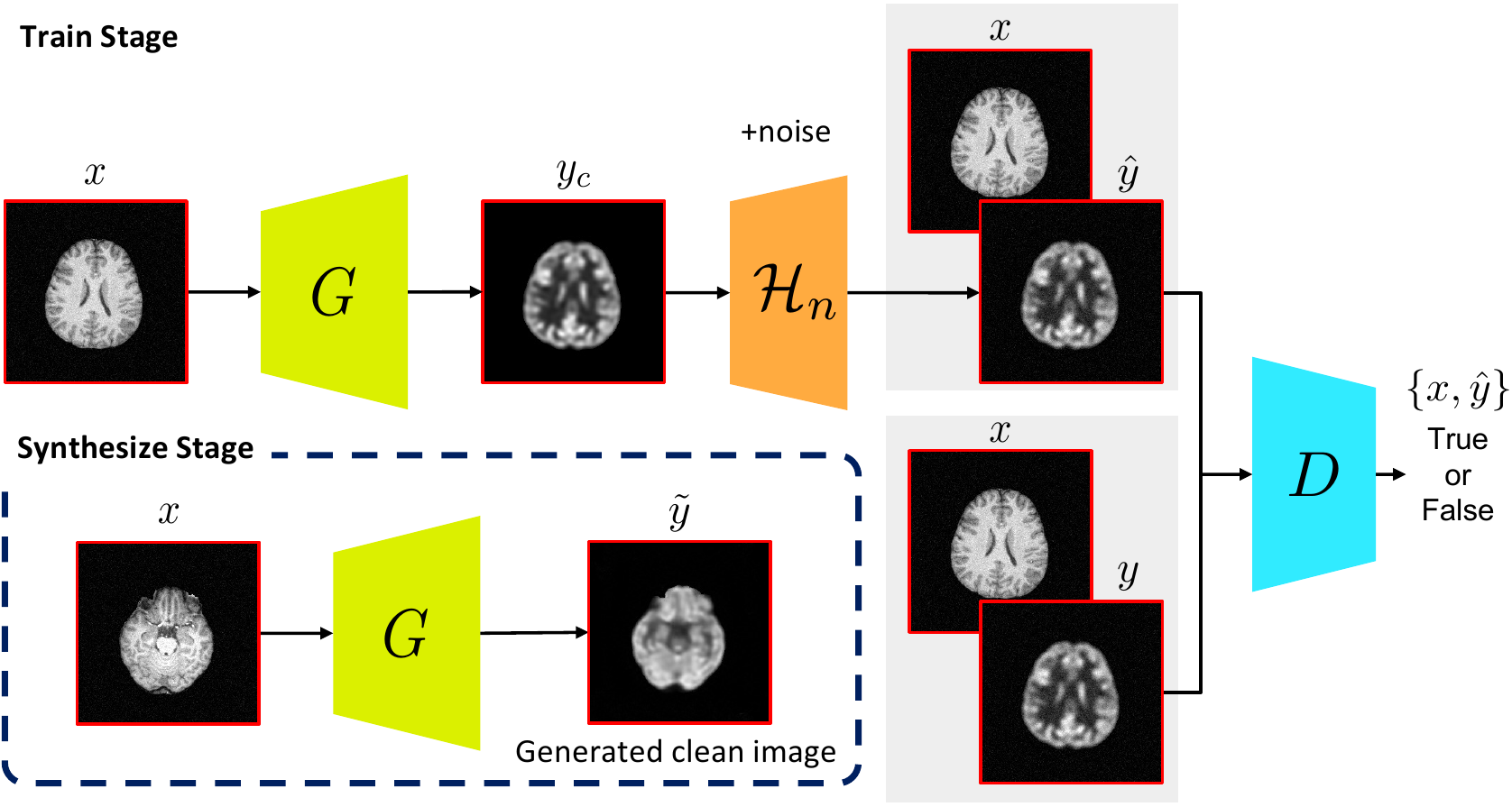}
    \caption{An illustration of the Ambient-Pix2PixGAN architecture in the image-to-image translation task from MRI ($x$) to PET ($y$). The generator $G$ is trained to generate a fake image $y_c$ that is processed by the image degradation function $\mathcal{H}_n$ to generate the fake measurement data $\hat{y}$ to fool $D$. The discriminator $D$ is trained to classify between fake (synthesized by the generator) and real tuples $\{x, \hat{y}\}$ and $\{x, y\}$.}
    \label{illustration}
\end{figure}
The objective function of the Ambient-Pix2PixGAN can be expressed as:
\begin{equation}
    \mathcal{L}=\mathcal{L}_{cGAN}+\lambda \mathcal{L}_{L1}.
    \label{all_loss}
\end{equation}
Here, $\mathcal{L}_{cGAN}$ is the adversarial loss that involves the simulated measurement data and the training measurement data, $\mathcal{L}_{L1}$ is the $L1$ distance between the fake measurement data and the ground-truth measurement data, and $\lambda$ is a parameter that controls the relative importance between the adversarial loss and the $L1$ distance.
The $\mathcal{L}_{cGAN}$ and $\mathcal{L}_{L1}$ can be expressed as:
\begin{equation}
\mathcal{L}_{cGAN}= \mathbb{E}_{x, \hat{y}}[\log D(x, \hat{y})]+\mathbb{E}_{x}[\log (1-D(x, y)],
\end{equation}
\begin{equation}
    \mathcal{L}_{L1}=\mathbb{E}_{y, \hat{y}}\left[\|\hat{y}-y\|_1\right].
\end{equation}
Here, $G$ and $D$ represent the generator network and discriminator network, respectively, $x$ denotes the training data in the source domain and $y$ denotes the training data in the target domain, $y_{c}$ is 
the generator-produced image: $y_{c}=G(x)$, and $\hat{y}$ is the fake measurement data $\hat{y}=\mathcal{H}_n({y_c})$.

\section{NUMERICAL STUDIES AND RESULTS}
A total number of 4,503 2D brain MRI and PET images from 36 healthy subjects were used in this work. The data had been de-identified and approved by the Department of Nuclear Medicine, Ruijin Hospital, Shanghai Jiao Tong University School of Medicine. In this study, 3,484 (80\%) subjects were used for training and 1,019 (20\%) were used for testing. The data had the size of $256 \times 256$. To simulate the noisy measured images, we added Gaussian noise with a mean of 0 and a standard deviation of 0.05 to the MRI and PET images. Our experiments were conducted on one NVIDIA GeForce RTX 3090 Ti GPU with the initial learning rate of model training set to 2e-4 and the batch size to 4. The total number of iterations is 2000. We used minibatch SGD and apply the Adam solver, with a learning rate of 0.0002, and momentum parameters $\beta_1=0.5$, $\beta_2=0.999$. During training, we also set the hyper-parameter $\lambda=1$ in the objective loss function $\mathcal{L}$ in Eq. \ref{all_loss}.

\begin{figure}[!ht]
\centering
\subfigure[Pix2PixGAN\label{pet_pix2pix}]{
    \begin{minipage}[t]{0.48\linewidth}
 	\centering
 	\includegraphics[width=\textwidth]{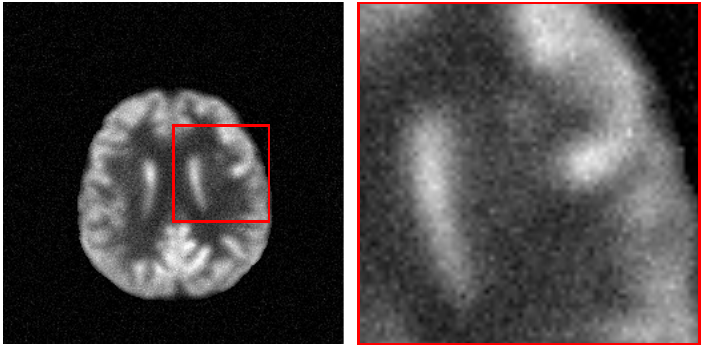}\\
 	\end{minipage}
}
\subfigure[Ambient-Pix2PixGAN\label{pet_ambient}]{
    \begin{minipage}[t]{0.48\linewidth}
 	\centering
 	\includegraphics[width=\textwidth]{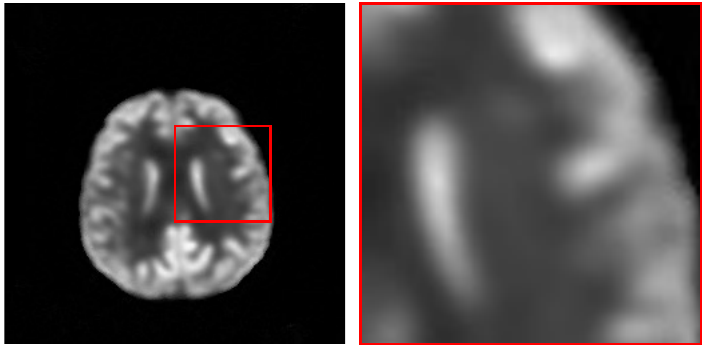}\\
 	\end{minipage}
}
\centering
\caption{(a) and (b) are images generated by the Pix2PixGAN and Ambient-Pix2PixGAN, respectively.}
\label{img_comp}
\end{figure}

Examples of images produced by the Ambient-Pix2PixGAN are shown in Fig. \ref{illustration}. The input MRI images are noisy while the synthesized PET images are clean.
\begin{figure}[!ht]
    \centering
    \includegraphics[width=\textwidth]{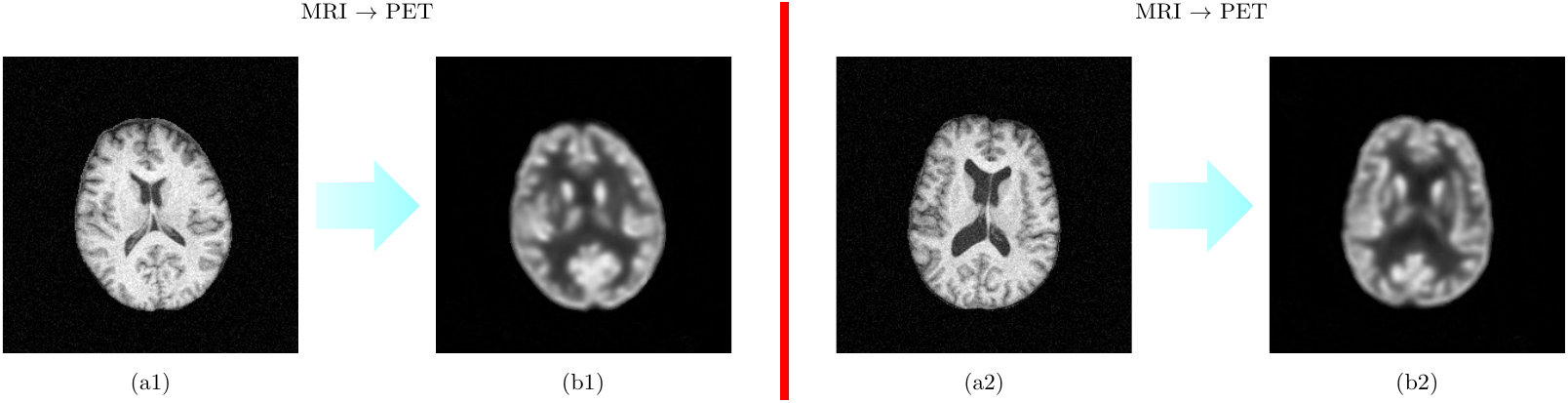}
    \caption{(a1) and (a2) are the noisy MRI images as input, (b1) and (b2) are the corresponding PET images generated by our proposed Ambient-Pix2PixGAN.}
    \label{illustration}
\end{figure}
Additionally, a traditional Pix2PixGAN was trained directly on noisy MRI and PET images. The generated images were compared with those produced by the Ambient-Pix2PixGAN in Fig. \ref{img_comp}. As expected, the traditional Pix2PixGAN that directly trained on noisy images produced synthesized images that are strongly affected by noise; while the images produced by the proposed Ambient-Pix2PixGAN are clean because the Ambient-Pix2PixGAN was trained to learn the underlying distribution of objects instead of noisy measurement data.

\begin{figure}[!ht]
\centering
\subfigure[Singular value spectrum\label{svs}]{
    \begin{minipage}[t]{0.48\linewidth}
 	\centering
 	\includegraphics[width=\textwidth]{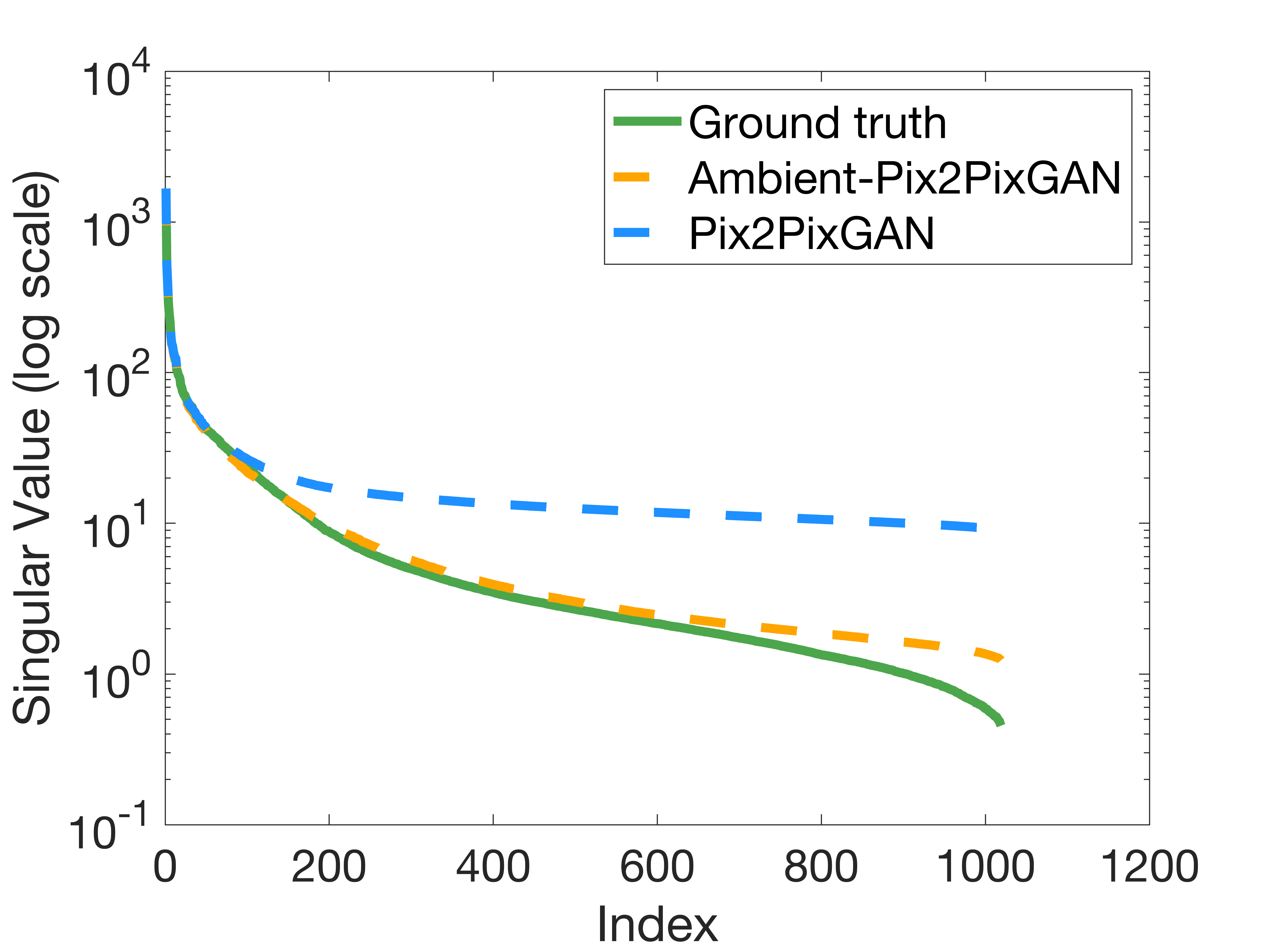}\\
 	\end{minipage}
}
\subfigure[Power spectrum\label{ps}]{
    \begin{minipage}[t]{0.48\linewidth}
 	\centering
 	\includegraphics[width=\textwidth]{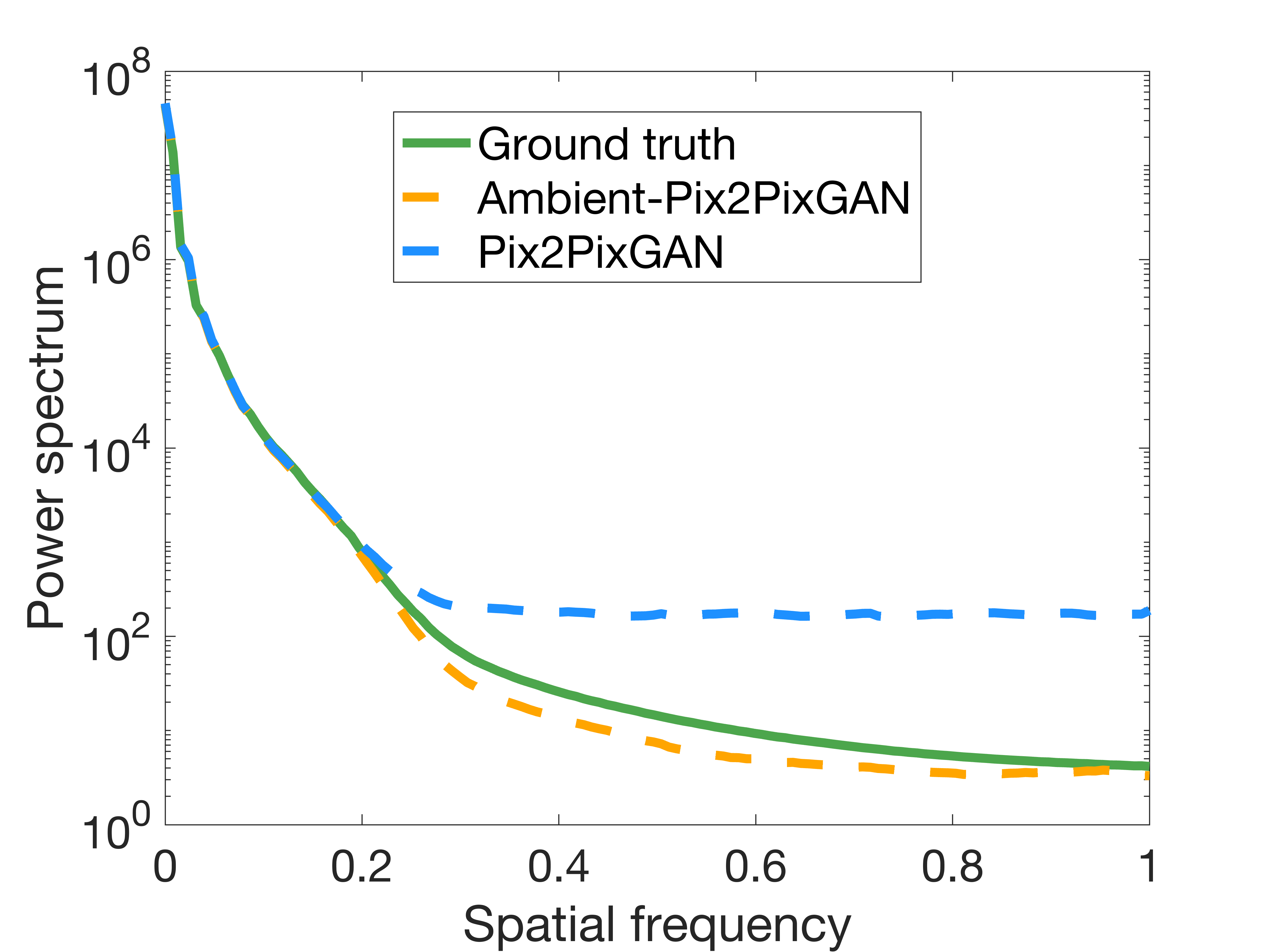}\\
 	\end{minipage}
}\\
\subfigure[SSIM\label{vis_ssim}]{
    \begin{minipage}[t]{0.31\linewidth}
 	\centering
 	\includegraphics[width=\textwidth]{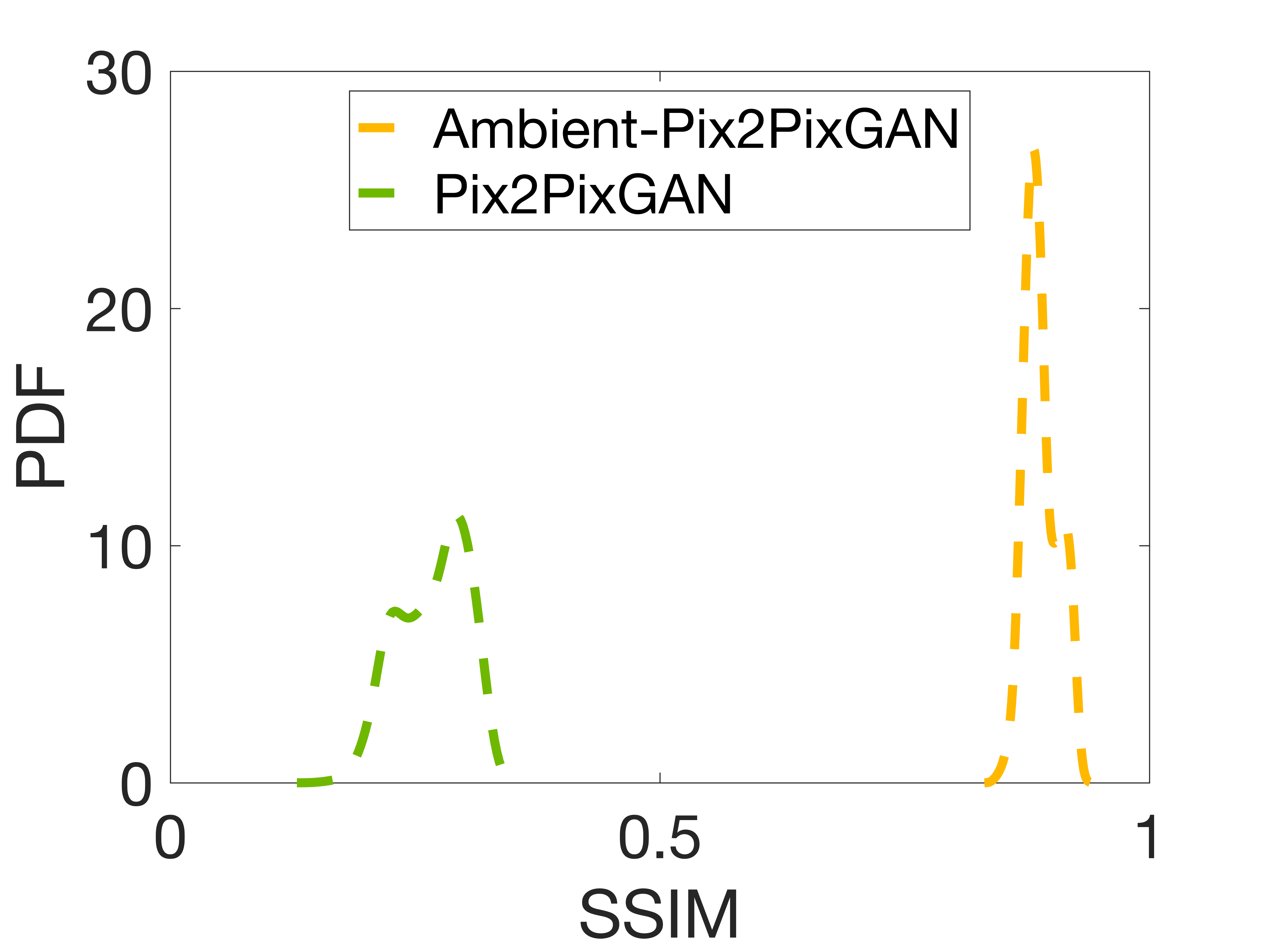}\\
 	\end{minipage}
}
\subfigure[PSNR\label{vis_psnr}]{
    \begin{minipage}[t]{0.31\linewidth}
 	\centering
 	\includegraphics[width=\textwidth]{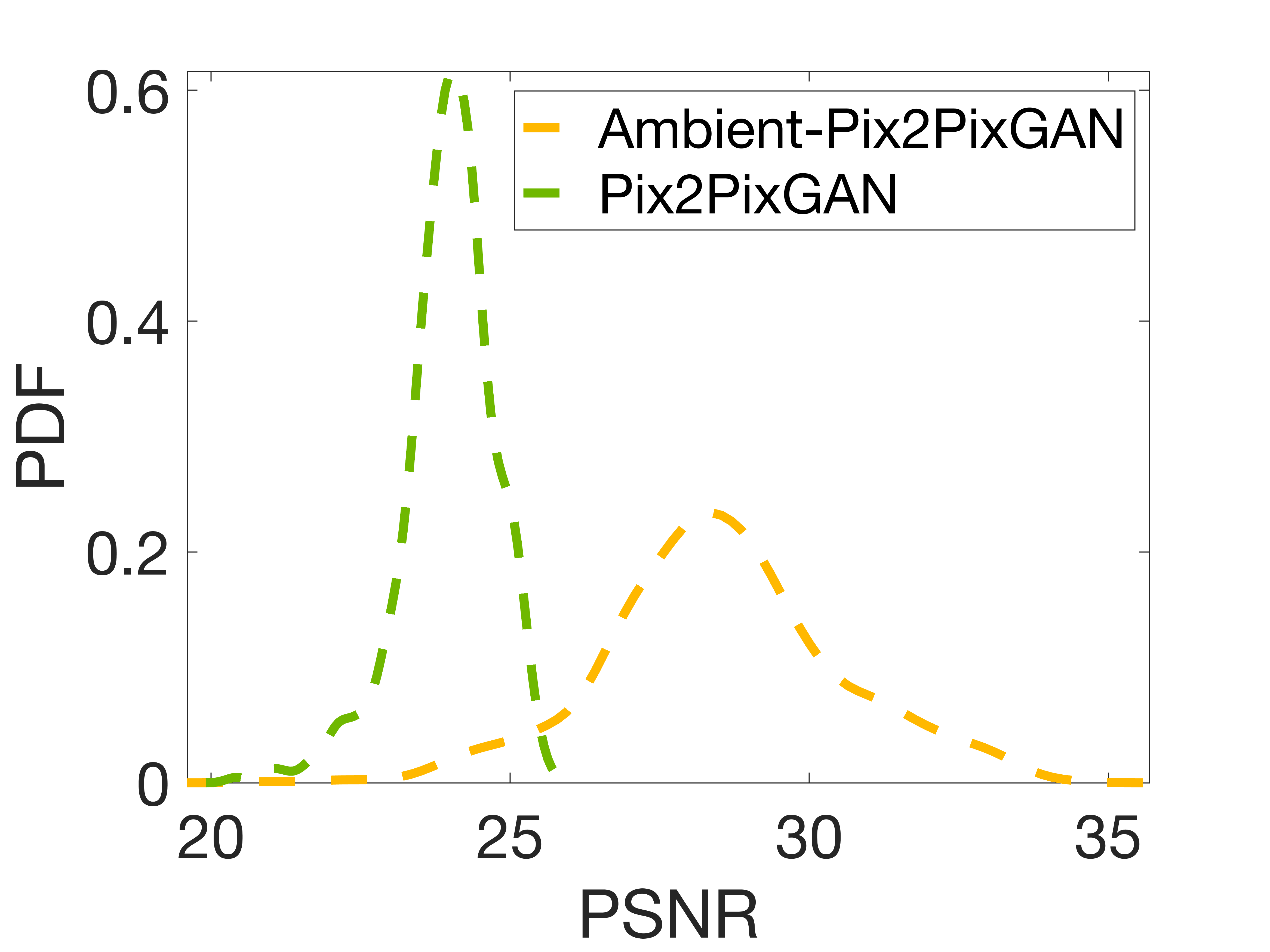}\\
 	\end{minipage}
}
\subfigure[RMSE\label{vis_rmse}]{
    \begin{minipage}[t]{0.31\linewidth}
 	\centering
        \renewcommand{\thefigure}{a1}
 	\includegraphics[width=\textwidth]{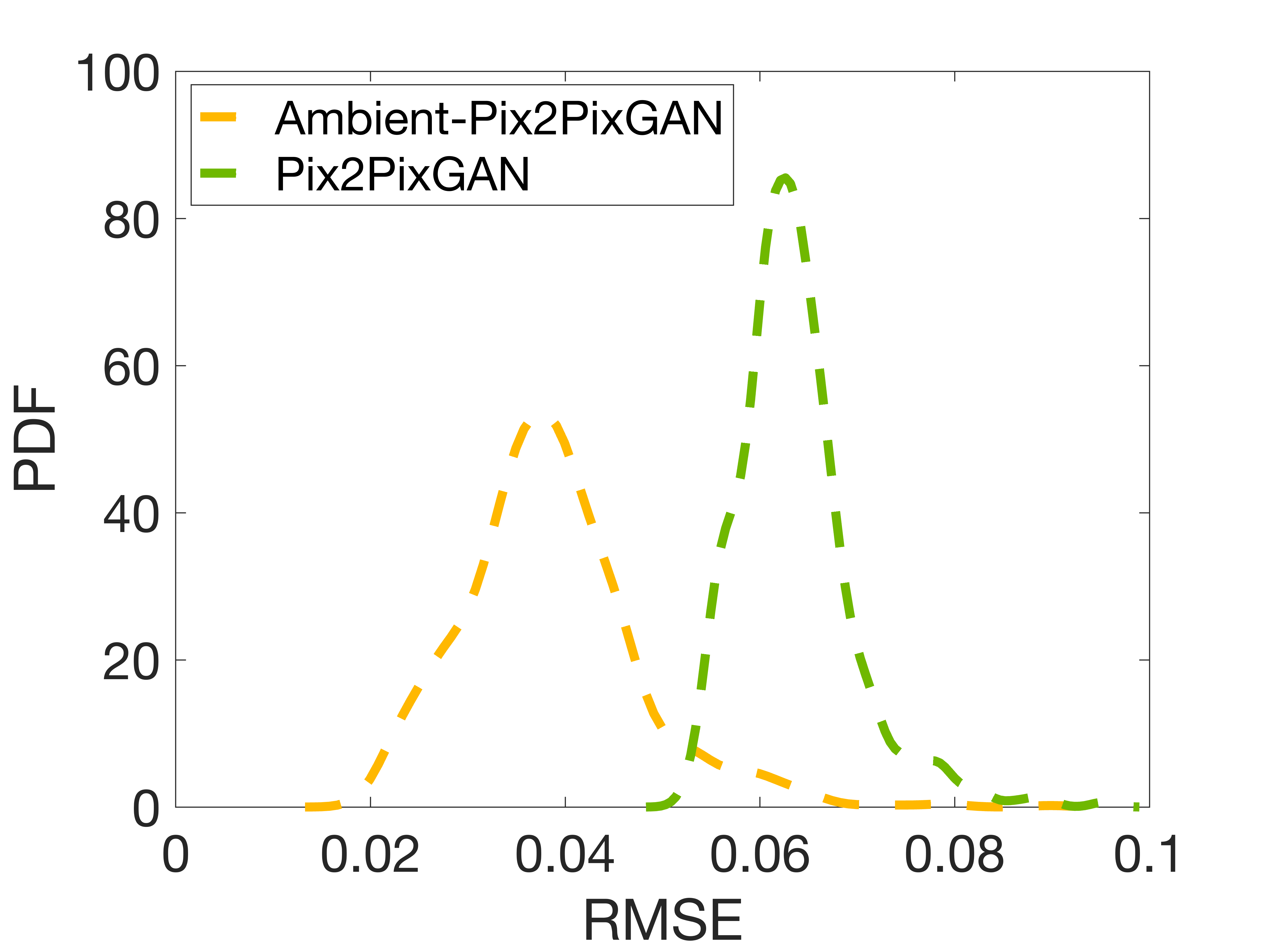}\\
 	\end{minipage}
}
\centering
\caption{(a) Singular value spectra comparison, (b) Radially averaged power spectra comparison, (c)-(e) represent probability density functions (PDFs) corresponding to SSIM, PSNR, and RMSE, respectively.}
\label{vis}
\end{figure}

The Ambient-Pix2PixGAN and the traditional Pix2PixGAN were further evaluated by use of Structure Similarity Index Measure (SSIM), Frechet Inception Distance (FID), Peak Signal-to-Noise Ratio (PSNR) and Root Mean Squared Error (RMSE). These metrics were evaluated on all 1,019 images in the test set, and the results are shown in Table \ref{results}. Compared with the traditional Pix2PixGAN, the proposed Ambient-Pix2PixGAN achieved improved values for all the considered metrics. We also calculated the singular value spectrum of the covariance matrix for images generated by Pix2PixGAN and Ambient-Pix2PixGAN, as well as for the ground truth images. The three singular value spectra are compared in Fig. \ref{svs}. It is shown that the curve corresponding to the Ambient-Pix2PixGAN has a higher degree of overlap with the ground truth curve compared to that corresponding to the Pix2PixGAN. We also compared the radially averaged power spectrum corresponding to the generated images and ground truth images in Fig. \ref{ps}. Again, the power spectrum corresponding to the Ambient-Pix2PixGAN is closer to the ground-truth power spectrum than the traditional Pix2PixGAN.

\begin{table}[!ht]
\centering
\begin{tabular}{cccccc}
\toprule
Model              & Noise                     & SSIM ($\uparrow$) & FID ($\downarrow$) & PSNR ($\uparrow$) & RMSE ($\downarrow$) \\ \midrule
Pix2PixGAN         & \multirow{2}{*}{std=0.05} & 0.2677 & 87.9604 & 23.9667 & 0.0636      \\
Ambient-Pix2PixGAN &                           & 0.8892 & 47.0719 & 28.5325 & 0.0384     \\  \bottomrule
\end{tabular}
\caption{Performance comparison between our method (Ambient-Pix2PixGAN) and Pix2PixGAN.}
\label{results}
\end{table}

Additionally, probability density functions (PDFs) of SSIM, PSNR, and RMSE evaluated between synthesized images and ground-truth images are shown in Fig \ref{vis_ssim} to Fig. \ref{vis_rmse}. It is shown that the SSIM values and PSNR vales produced by the Ambient-Pix2PixGAN were overall greater than those produced by the Pix2PixGAN, and the RMSE values produced by the Ambient-Pix2PixGAN were overall smaller than those produced by the Pix2PixGAN.

Moreover, a task-based image quality evaluation was conducted to assess our proposed Ambient-Pix2PixGAN method. Specifically, we considered a signal-known-exactly (SKE) and background-known-statistically (BKS) binary signal detection task. An observer is required to classify the measured image data $\mathbf{g}$ as satisfying either a signal-present hypothesis $H_{1}$ or a signal-absent hypothesis $H_{0}$ \cite{barrett2013foundations}. The imaging processes under these two hypotheses can be described as:
\begin{subequations}
\label{eq:imgH_s}
\begin{align}
    H_{0}:&\ \mathbf{g}=\mathbf{b}+\mathbf{n},\\
    H_{1}:&\ \mathbf{g}=\mathbf{b}+\mathbf{s}+\mathbf{n}.
\end{align}
\end{subequations}
The background $\mathbf{b}$ can be either the ground-truth PET images or the generated images. The PET images were center-cropped to a size of $30 \times 30$ and a Gaussian signal was added to form the signal-present images. Table \ref{snrho} shows the specifications of the five considered signal detection tasks.

\begin{figure}[!ht]
    \centering
    \includegraphics[width=0.9\textwidth]{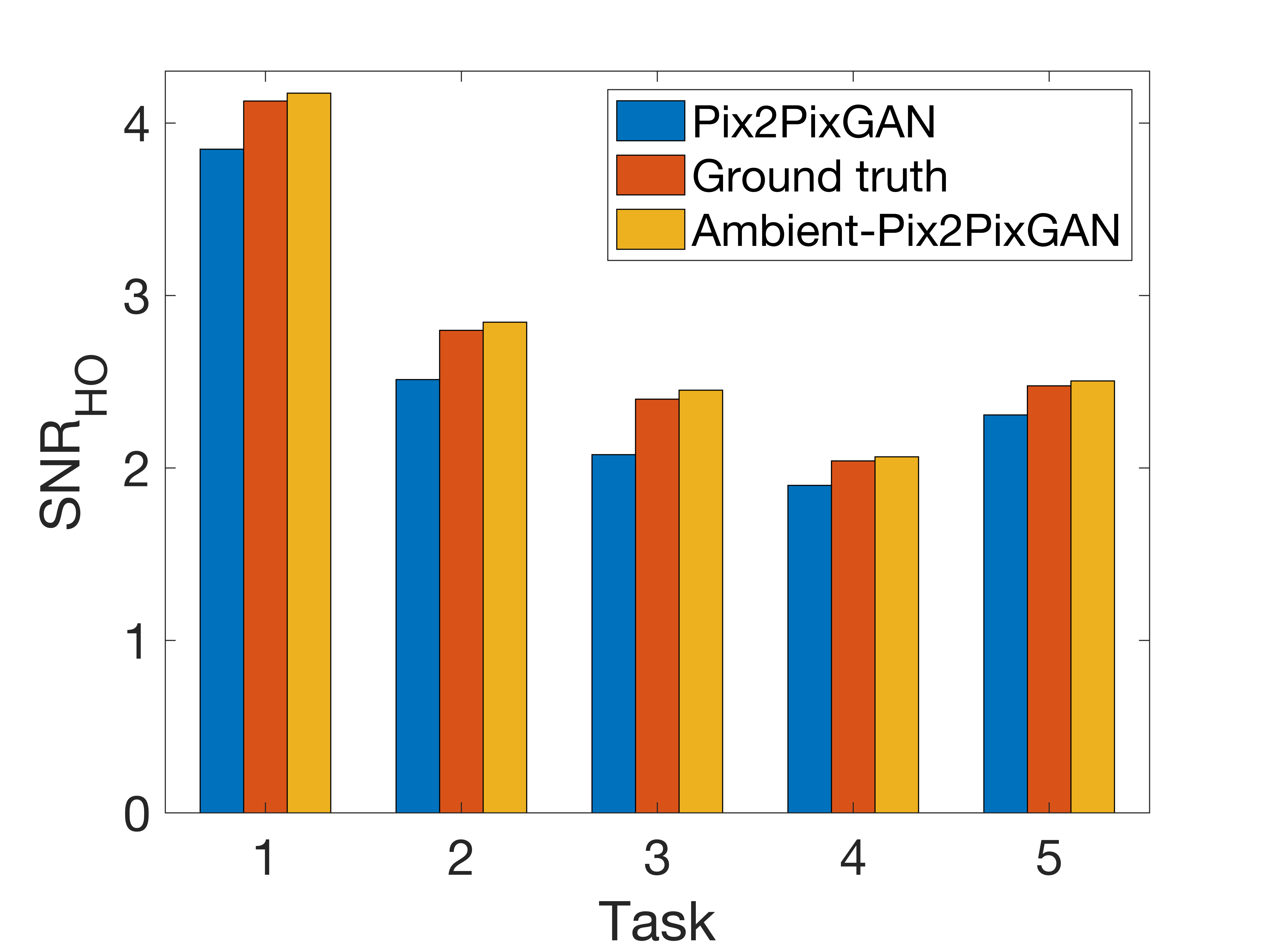}
    \caption{Comparison between Pix2PixGAN generated, Ambient-Pix2PixGAN generated and ground truth $\text{SNR}_{HO}$ for different signal detection tasks.}
    \label{snr_ho}
\end{figure}

\begin{table}[!ht]
\centering
\begin{tabular}{ccc}
\toprule
Task & Signal & Noise \\ \midrule
1    &   Gaussian Signal(std=0.6, magnitude=1.5)   &  Gaussian Noise(mean=0, std=1)       \\
2     &  Gaussian Signal(std=0.5, magnitude=1)   &  Gaussian Noise(mean=0, std=0.7)      \\
3     &  Gaussian Signal(std=0.7, magnitude=0.4)   &  Gaussian Noise(mean=0, std=0.5)       \\
4     &  Gaussian Signal(std=0.5, magnitude=1)   &  Gaussian Noise(mean=0, std=1)       \\
5     &  Gaussian Signal(std=0.6, magnitude=0.9)   &  Gaussian Noise(mean=0, std=1)       \\ \bottomrule
\end{tabular}
\caption{Five considered signal detection tasks, the location of the added signal is the center of the image.}
\label{snrho}
\end{table}
The signal-to-noise ratio of the test statistic of the Hotelling observer, $\text{SNR}_{HO}$, was employed to assess the signal detection performance. The $\text{SNR}_{HO}$ can be computed as \cite{barrett2013foundations}:
\begin{equation}
\mathrm{SNR}_{HO}^2=\Delta\overline{\overline{\mathbf{g}}}^T\mathbf{w}_{\mathrm{HO}}=\Delta\overline{\overline{\mathbf{g}}}^T\left[\mathbf{K}_{\mathbf{g}}\right]^{-1} \Delta \overline{\overline{\mathrm{g}}}.
\end{equation}
where $\Delta \overline{\overline{\mathrm{g}}}$ is the difference between means of vector $\mathbf{g}$ under the $H_1$ and $H_0$ hypotheses, and $\mathbf{K}_{\mathbf{g}}$ is the covariance matrix of the measured data $\mathbf{g}$. The $\text{SNR}_{HO}$ values produced by the Pix2PixGAN, Ambient-Pix2PixGAN, and the ground truth images are shown in Fig. \ref{snr_ho}. The Pix2PixGAN produced smaller $\text{SNR}_{HO}$ because it produced images that are affected by noise; while the Ambient-Pix2PixGAN produced $\text{SNR}_{HO}$ values similar to the ground truth values.

\section{Conclusion}
This study introduces a novel Ambient-Pix2PixGAN approach for performing paired image-to-image translation tasks for cases where only noisy image data are available. Preliminary studies that considered noisy MRI and PET images were conducted. Both traditional image quality measures and task-based image quality measures were evaluated to assess the trained models. It was demonstrated that Ambient-Pix2PixGAN can be successfully applied to synthesize high-quality clean images based on noisy data.

\bibliography{report} 
\bibliographystyle{spiebib} 

\end{document}